\documentclass[a4paper,12pt, epsfig]{article}
\usepackage{epsfig}
\usepackage{amssymb,latexsym}
\usepackage{amsfonts}
\usepackage{amsmath}
\usepackage{graphicx}
\newskip\humongous \humongous=0pt plus 1000pt minus 1000pt

\newif\ifdtup

\catcode`\@=11

\@addtoreset{equation}{section}
\def\theequation{\thesection.\arabic{equation}}

\def\@normalsize{\@setsize\normalsize{15pt}\xiipt\@xiipt
\abovedisplayskip 14pt plus3pt minus3pt%
\belowdisplayskip \abovedisplayskip
\abovedisplayshortskip \z@ plus3pt%
\belowdisplayshortskip 7pt plus3.5pt minus0pt}

\def\small{\@setsize\small{13.6pt}\xipt\@xipt
\abovedisplayskip 13pt plus3pt minus3pt%
\belowdisplayskip \abovedisplayskip
\abovedisplayshortskip \z@ plus3pt%
\belowdisplayshortskip 7pt plus3.5pt minus0pt
\def\@listi{\parsep 4.5pt plus 2pt minus 1pt
      \itemsep \parsep
      \topsep 9pt plus 3pt minus 3pt}}

\relax

\catcode`@=12

\catcode`\@=11

\def\section{\@startsection{section}{1}{\z@}{3.5ex plus 1ex minus
    .2ex}{2.3ex plus .2ex}{\large\bf}}

\def\thesection{\arabic{section}}
\def\thesubsection{\arabic{section}.\arabic{subsection}}

\def\appendix{\setcounter{section}{0}
  \def\thesection{Appendix \Alph{section}}
  \def\thesubsection{\Alph{section}.\arabic{subsection}}
  \def\theequation{\Alph{section}.\arabic{equation}}}
\def\SymBoxes#1#2#3#4{\newdimen\un@t \un@t#3%
\raisebox{#1}{\rule{#2\un@t}{#4}\hskip-#2\un@t% lower horizontal
\@tempdimb\un@t \advance\@tempdimb by-#4\@tempcntb#2\relax%
\@whilenum{\@tempcntb>0}\do{%                         % #2 vertical lines
\rule{#4}{\un@t}\hskip\@tempdimb \advance\@tempcntb by\m@ne}%
\hskip-#2\un@t \rule[\un@t]{#2\un@t}{#4}%
\rule[\un@t]{#4}{#4}\hskip-#4%             % upper horizontal line
\rule{#4}{\un@t}}\hskip-#4}                % rightest vertical line
\begin{document}
%\begin{letter}{~}

%%%%%%Define some new commands and  macros

\newcommand{\dd}{\textrm{d}}

\newcommand{\beq}{\begin{equation}}
\newcommand{\eeq}{\end{equation}}
\newcommand{\bea}{\begin{eqnarray}}
\newcommand{\eea}{\end{eqnarray}}
\newcommand{\beas}{\begin{eqnarray*}}
\newcommand{\eeas}{\end{eqnarray*}}
\newcommand{\defi}{\stackrel{\rm def}{=}}
\newcommand{\non}{\nonumber}
\newcommand{\bquo}{\begin{quote}}
\newcommand{\enqu}{\end{quote}}
%%%%%%%%%%%%%%%%
\renewcommand{\(}{\begin{equation}}
\renewcommand{\)}{\end{equation}}
%%%%%%%%%%%%%%%%%%%%%%%%%%%%%%%%%% definitions
\def\de{\partial}
\def\Om{\ensuremath{\Omega}}
\def\Tr{ \hbox{\rm Tr}}
\def\rc{ \hbox{$r_{\rm c}$}}
\def\H{ \hbox{\rm H}}
\def\HE{ \hbox{$\rm H^{even}$}}
\def\HO{ \hbox{$\rm H^{odd}$}}
\def\HEO{ \hbox{$\rm H^{even/odd}$}}
\def\HOE{ \hbox{$\rm H^{odd/even}$}}
\def\HHO{ \hbox{$\rm H_H^{odd}$}}
\def\HHEO{ \hbox{$\rm H_H^{even/odd}$}}
\def\HHOE{ \hbox{$\rm H_H^{odd/even}$}}
\def\K{ \hbox{\rm K}}
\def\Im{ \hbox{\rm Im}}
\def\Ker{ \hbox{\rm Ker}}
\def\const{\hbox {\rm const.}}
\def\o{\over}
\def\im{\hbox{\rm Im}}
\def\re{\hbox{\rm Re}}
\def\bra{\langle}\def\ket{\rangle}
\def\Arg{\hbox {\rm Arg}}
\def\exo{\hbox {\rm exp}}
\def\diag{\hbox{\rm diag}}
\def\longvert{{\rule[-2mm]{0.1mm}{7mm}}\,}
\def\a{\alpha}
\def\b{\beta}
\def\e{\epsilon}
\def\l{\lambda}
\def\ol{{\overline{\lambda}}}
\def\ochi{{\overline{\chi}}}
\def\th{\theta}
\def\s{\sigma}
\def\oth{\overline{\theta}}
\def\ad{{\dot{\alpha}}}
\def\bd{{\dot{\beta}}}
\def\oD{\overline{D}}
\def\opsi{\overline{\psi}}
\def\dag{{}^{\dagger}}
\def\tq{{\widetilde q}}
\def\L{{\mathcal{L}}}
\def\p{{}^{\prime}}
\def\W{W}
\def\N{{\cal N}}
\def\hsp{,\hspace{.7cm}}
\def\bo{\ensuremath{\hat{b}_1}}
\def\bfo{\ensuremath{\hat{b}_4}}
\def\co{\ensuremath{\hat{c}_1}}
\def\cfo{\ensuremath{\hat{c}_4}}
\newcommand{\C}{\ensuremath{\mathbb C}}
\newcommand{\Z}{\ensuremath{\mathbb Z}}
\newcommand{\R}{\ensuremath{\mathbb R}}
\newcommand{\rp}{\ensuremath{\mathbb {RP}}}
\newcommand{\cp}{\ensuremath{\mathbb {CP}}}
\newcommand{\vac}{\ensuremath{|0\rangle}}
\newcommand{\vact}{\ensuremath{|00\rangle}                    }
\newcommand{\oc}{\ensuremath{\overline{c}}}
\newcommand{\Cos}{\textrm{cos}}

\newcommand{\Vol}{\textrm{Vol}}

\newcommand{\half}{\frac{1}{2}}

%%%%%%%%%%%%%%%%%%%%%%%Changed%%%%%%%%%%%%%%%%%%%%%%%%%%%%%
\def\changed#1{{\bf #1}}
%\def\changed#1{ #1}
%%%%%%%%%%%%%%%%%%%%%%%%%%%%%%%%%%%%%%%%%%%%%%%%%%%%%%%%%

\begin{titlepage}
%\begin{flushright}
%IFUP-TH/2011-??
%\end{flushright}
%\bigskip

\def\thefootnote{\fnsymbol{footnote}}

\begin{center}
{\large {\bf
Neutrino Splitting and Density-Dependent Dispersion Relations 
  } }
%\end{center}

\bigskip

\bigskip

{\large \noindent Emilio
Ciuffoli$^{1}$\footnote{ciuffoli@ihep.ac.cn}, Jarah
Evslin$^{1}$\footnote{\texttt{jarah@ihep.ac.cn}}, Xiaojun Bi${}^{2}$\footnote{\texttt{bixj@ihep.ac.cn}}  
and Xinmin Zhang$^{3,1}$\footnote{\texttt{xmzhang@ihep.ac.cn}} }
\end{center}

\renewcommand{\thefootnote}{\arabic{footnote}}

\vskip.7cm

\begin{center}
\vspace{0em} {\em  { 1) TPCSF, IHEP, Chinese Acad. of Sciences\\
2) Key laboratory of particle astrophysics, IHEP, Chinese
Acad. of Sciences,\\
3) Theoretical physics division, IHEP, Chinese Acad. of Sciences\\
YuQuan Lu 19(B), Beijing 100049, China}}

\vskip .4cm

\end{center}

\vspace{4.5cm}

\noindent
\begin{center} {\bf Abstract} \end{center}

\noindent 
We show that particles can split only when their group velocity exceeds their phase velocity.  In this sense the splitting process is the quantum analog of the modulational instability in anomalous dispersive media.  In the case of a neutrino whose dispersion relation contains a subdominant Lorentz-violating correction of the form $aP^k$, the neutrino will decay into two neutrinos and an antineutrino at a rate proportional to $a^3G_F^2E^{2+3k}$.  Unlike the Cohen-Glashow instability, the splitting instability exists even if all particles involved in the interaction have the same dispersion relations at the relevant energy scales.  We show that this instability leads to strong constraints even if the energy $E$ is a function of both the momentum $P$ and also of the background density $\rho$, for example we show that it alone would have been sufficient to eliminate any model of the MINOS/OPERA velocity anomaly which modifies the neutrino dispersion relation while leaving those of other particles intact.

%We consider modifications of the Standard Model in which {\it{only}} the neutrino dispersion relation is changed, allowing the energy $E$ to be a function of the momentum $P$ and also of the background density $\rho$.  The $\rho$ dependence can easily accommodate SN1987A constraints.  By considering only experiments in the Earth's crust, $\rho$ is fixed, leaving a function $E(P)$.  Assuming a weak monotonicity constraint and using bounds from $e^-e^+$  emission (Bremsstrahlung) and $\nu\bar{\nu}$ emission (neutrino splitting), we show that no function is simultaneously compatible with (1) the neutrino energy distribution observed at OPERA, (2) the neutrino velocity reported at OPERA and (3) the absence of high energy $e^-e^+$ pairs at ICARUS.  A key role is played by a bound on the concavity of the dispersion relation imposed by neutrino splitting, a technique which is argued to have several other applications.  Our results do not exclude models which include new physics in addition to a modified neutrino dispersion relation.

\vfill

\begin{flushleft}
{\today}
%\vspace{1cm}
\end{flushleft}
\end{titlepage}
%\bigskip

\hfill{}
%\bigskip

%\tableofcontents

\setcounter{footnote}{0}
\def\ie{\emph{i.e. }}
\def\ds{\partial\!\!\!/}

%%%%%%%%%%%%%%%%%%%%%%%Changed%%%%%%%%%%%%%%%%%%%%%%%%%%%%%
\def\changed#1{{\bf #1}}
%%%%%%%%%%%%%%%%%%%%%%%%%%%%%%%%%%%%%%%%%%%%%%%%%%%%
%\section{Introduction}
\section{Motivation}
The velocity anomaly reported by OPERA \cite{opera} was the result of experimental error \cite{zichichi}.  The frenzied activity following OPERA's announcement is often criticized for its inconsistent attempts at profound advances in fundamental physics.  However it also led to several noteable advances of general interest, in particular with regard to constraints on unusual dispersion relations. The most celebrated among these is that of Cohen and Glashow \cite{cg} who demonstrated that if at high energies the velocities of two species of particle asymptote to different values, then the fast particle will lose energy as it travels, transferring it into a succession of slow particles.  Similarly it was demonstrated that the phase space for a slow particle to decay into fast particles is extremely limited \cite{xiaojun}.  

These two new constraints were powerful enough to eliminate all of the models which had been formulated at that time, even those with exotic dispersion relations such as those of Refs. \cite{ealfa,strumia}.  For example, consider the model of Ref. \cite{ealfa} in which the neutrino velocity is a power $P^k$ of the momentum $P$, with $k>0$.  As we will explain, when $k>0.5$ these models were ruled experimentally by OPERA's short burst run \cite{opera} in which the energy varied by at least a factor of 5 but the velocity varied by less than a factor of 2.  On the other hand $k>2.5$ is required for consistency with supernova bounds \cite{snneut,snphot}.  These models can also be ruled out theoretically, as is easily seen by extending Cohen and Glashow's calculation to the proposed dispersion relation.  Such generalizations were considered in Refs. \cite{miao,dispersione,piudispersione}, where it was seen that they only change the decay rate by a $k$-dependent geometrical factor of order 1, not by nearly enough to avoid the Cohen-Glashow electron-positron nucleation process.  As a result the neutrino energy decays appreciably before arriving at OPERA, in contradiction with the fact that the observed spectrum agreed well with simulations up to 100 GeV \cite{operafeb}.  

The model of Ref \cite{strumia} is consistent with the results of the short burst run, as the velocity is constant throughout the energy range probed.  However the dispersion relation is nearly identical to that of Cohen and Glashow, with the only difference arising in a small portion of phase space at small momenta.  Thus again an application of Cohen and Glashow's argument implies that in this model the neutrinos should have lost most of their energy before arriving at the OPERA detector, in contradiction with observations.

Refs. \cite{cg,xiaojun} remain relevant even now that the velocity anomaly has disappeared because they provide general constraints on dispersion relations that differ from those of special relativity.  As is clearly explained in the first two paragraphs of Ref. \cite{einstein}, the dispersion relation of special relativity only applies to particles in the vacuum.  In particular it applies to particles in a Lorentz-invariant vacuum.  It therefore does not apply to many systems of interest, such as neutrinos traveling through the Earth or particles in the primordial plasma.  If the acceleration of the universe is caused by an interacting dark energy field, then the expectation value of this field is not Lorentz-invariant, it depends on a universal time.  As a result, at some scale the dispersion relations of all particles will deviate from the special relativistic form.  Even if dark energy is simply caused by a cosmological constant, the FRLW solution of our universe is not Lorentz-invariant, and so one may expect a combination of Hubble and Planck scale corrections to the relativistic dispersion relations.  

More immediately, neutrinos traveling inside of media still exhibit a number of anomalies.  For example there is a solar neutrino deficit at low energies \cite{smirnov} and both AMANDA and IceCube observe a surplus of horizontal neutrinos in ice \cite{icecube}.  LSND, whose baseline is dominated by Earth and iron shielding has seen an anomalous neutrino deficit \cite{lsnd} while KARMEN, whose baseline is dominated by air, has not \cite{karmen}.  Needless to say, these media are not Lorentz-invariant and so the neutrino dispersion relations inside of them, even within the context of the standard model \cite{msw}, are not Lorentz-invariant.  While it is not yet clear whether these anomalies are due to new physics, they motivate a better understanding of constraints on neutrino dispersion relations in general and in particular of neutrino dispersion relations in media.

While the constraints \cite{cg,xiaojun} are quite powerful, eventually models of superluminal neutrinos were created which avoided the assumptions implicit in these works and so avoided the constraints. The most obvious assumption is 4-momentum conservation, and discussions of escapes along those lines first appeared in Refs.~\cite{momcon}.  It was also argued that one could escape by rendering other particles, such as the electron, superluminal \cite{me,tedeschi} after all the dispersion relations could well be density dependent \cite{pospelov,ether,tianjun} and all strong constraints on electron velocities appear to come from experiments in a vacuum \cite{tianjun}.  Another escape route is if, while traveling, the neutrinos are noninteracting because they are taking a short cut through another dimension \cite{xdim}, or converted into a sterile flavor \cite{sterile}.  While such models are no longer necessary to explain OPERA's anomaly, they may well resurface in attempts to explain the yet unexplained neutrino anomalies.  This leads one to ask if further constraints are available which may eliminate some of these models.

\section{Neutrino splitting} \label{splitsez}
\subsection{What is neutrino splitting?}

In this note we will argue that neutrino splitting \cite{splitvec,splitnuov}
\beq
\nu\longrightarrow 2\nu+ \overline{\nu} \label{proc}
\eeq
provides a powerful constraint.  Like the Cohen-Glashow and phase space constraints, the applications of this constraint extend far beyond OPERA.  The Cohen-Glashow process is only allowed if two different kinds of particles have different dispersion relations.  Therefore it is avoided in models with universal high energy dispersion relations such as \cite{tianjun,me,tedeschi}.  While qualitative features of such models were discussed in Ref. \cite{me}, a concrete model of this kind was proposed in Ref \cite{tedeschi}, considering interface conditions between regions with different densities and demonstrating that fifth force constraints can be satisfied in those regions.  The phase space constraint \cite{tianjun} uses the kinematics of the creation of the neutrinos to limit its propagation along the baseline.  However these neutrinos are created in a vacuum whereas the baseline is almost entirely solid rock, therefore even a small amount of environmental dependence is sufficient to eliminate these constraints.  In other words, the constraint of Refs. \cite{xiaojun} are satisfied in any theory in which the neutrinos are subluminal in a vacuum, such as those of Refs \cite{me,tedeschi}.

The neutrino splitting constraint is more robust.  It depends only on the dispersion relations of the neutrinos themselves, so it cannot be avoided by modifying the properties of the other particles.  Furthermore the splitting process can occur anywhere along the baseline, so unlike the phase space constraint of \cite{xiaojun} which only constrains the dispersion relations at the point where the neutrinos are created, the splitting constraint constrains the dispersion relations everywhere that the neutrinos travel and so in the region in which the velocity anomaly is claimed.

When can splitting occur?  Consider a field $\phi$.  Whenever the low energy effective theory has an effective vertex of the form $\phi^k$ with $k>2$ there is a process in which a quantum of the field $\phi$ can decay into $k-1$ quanta of $\phi$.  If the field carries a conserved charge\footnote{If the conserved charge generates a cyclic group of finite order $N$ then this remains true so long as $N$ does not divide $k$.} then charge conservation does not allow an effective vertex of the form $\phi^k$.  However, a vertex which includes $k$ copies of the field and $k$ of the antifield is allowed.  For example in the case of neutrinos the 4 neutrino Fermi vertex is allowed, which yields the desired process (\ref{proc}). 

\subsection{Splitting kinematics}

When is splitting kinematically allowed?  When there is a large phase space for splitting, particles do not propagate very far before losing all of their energy.  To obtain the strongest possible bounds one is interested in the regime where splitting is near threshold.  In this case a particle with an initial momentum ${\mathbf{P^{(0)}}}$ in the $x$ direction will decay into $k$ particles with momenta ${\mathbf{P^{(i)}}}$ which are nearly in the $x$ direction.  More concretely, if for simplicity we assume that the $x$ direction is a principal direction of our dispersion relation, then the energy $E^{(0)}$ of the original particle and the energy $E^{(i)}$ of a given outgoing particle are approximately quadratic in the corresponding momenta
\beq
E^{(0)}=E(P^{(0)}_x)\hsp
E^{(i)}=E(P^{(i)}_x)+\alpha (P^{(i)}_y)^2 \label{quadap}
\eeq
where $\alpha$ is a strictly positive constant, $E(p)$ is a function and the $y$ axis is chosen such that ${\mathbf{P^{(i)}}}$  lies on the $x-y$ plane.

In this approximation, we will now demonstrate the following claim:

\noindent
{\textbf{Claim: }}{\textit{Particle splitting  is only kinematically allowed when the group velocity exceeds the phase velocity.  More precisely, it is only allowed for particles with $x$-momentum $P^{(0)}_x$ such that, at some $x$-momentum $p\leq P^{(0)}_x$,}
\beq
\frac{\partial E(p)}{\partial p}>\frac{E(p)}{p}.
\eeq

First, we note that Eq.~(\ref{quadap}) implies that the energy of the $i$th particle is greater than or equal to $E(P^{(i)}_x)$ with equality only when $ P^{(i)}_y=0$.  This equality is only satisfied on a measure zero portion of the phase space and so such decays do not contribute to the decay amplitude.  This reflects the familiar fact that if a particle can only decay into colinear particles then the instability is only marginal.  Therefore, observable decays only occur when the energy $E^{(i)}$ of each particle is strictly greater than the corresponding function $E(P^{(i)}_x)$
\beq
E^{(i)}>E(P^{(i)}_x). \label{ediff}
\eeq

Now we are ready to prove the claim by contradiction.  Assume that the group velocity is less than or equal to the phase velocity
\beq
\frac{\partial E(p)}{\partial p}\leq\frac{E(p)}{p}
\eeq
for all momenta $p\leq P^{(0)}_x$.  Dividing both sides by $E(p)$ one obtains
\beq
\frac{\partial {\mathrm{ln}}(E(p))}{\partial p}\leq\frac{1}{p}. \label{asurdo}
\eeq
Now one may calculate $E^{(i)}$ using the fundamental theorem of calculus and then Eq.~(\ref{asurdo})
\beq
{\mathrm{ln}}(E(P^{(i)}_x))={\mathrm{ln}}(E(P^{(0)}_x))-\int_{P^{(i)}_x}^{P^{(0)}_x}\frac{\partial {\mathrm{ln}}(E(p))}{\partial p} dp\geq {\mathrm{ln}}(E(P^{(0)}_x))-\int_{P^{(i)}_x}^{P^{(0)}_x}\frac{1}{p}dp={\mathrm{ln}}\left(\frac{P^{(i)}_x E(P^{(0)}_x)}{P^{(0)}_x}\right)
\eeq
which can be exponentiated to yield the second inequality in
\beq
\frac{E^{(i)}}{E^{(0)}}>\frac{E(P^{(i)}_x)}{E(P^{(0)}_x)}\geq \frac{P^{(i)}_x}{P^{(0)}_x}.
\eeq
The first inequality follows from Eqs.~(\ref{quadap}) and (\ref{ediff}).  Now summing over all of the final particles $i$, the rightmost and leftmost expressions yield unity.  Thus we have proven that $1>1$, a contradiction.  

This implies that the initial assumptions that there is finite phase space for a splitting process and that the group velocity never exceeds the phase velocity at $p<P^{(0)}_x$ are incompatible.  So splitting is only allowed at momenta greater than momenta at which the group velocity exceeds the phase velocity, at least when one is close enough to threshold that the nearly colinear approximation (\ref{quadap}) may be applied.

What about the converse, does a group velocity faster than the phase velocity imply that the splitting process is kinematically allowed?  If the group velocity is faster than the phase velocity at some momentum $P^{(0)}_x$, then at momenta $p$ infinitesimally below $P^{(0)}_x$, the energy $E(p)$ will be less that $p\frac{E^{(0)}}{P^{(0)}_x}$.  Let $P_m$ be the lowest such momentum, so that for $P_m<p<P^{(0)}_x$
\beq
E(p)<\frac{pE^{(0)}}{P^{(0)}_x}. \label{meno}
\eeq
Then splitting is kinematically allowed so long as all of the decay products have $p>P_m$.  The problem is that if $P_m>P^{(0)}_x/3$ then conservation of momentum does not allow all neutrinos to have momenta $p>P_m$, since the total momentum must be $P^{(0)}_x$.  

Therefore we learn that if (\ref{meno}) holds for all $p$ such that $P^{(0)}_x/3\leq p<P^{(0)}_x$ then the splitting is kinematically allowed.  However even if it does not then splitting reactions of the kind
\beq
a\nu\longrightarrow (a+b)\nu+b\overline{\nu}
\eeq
are allowed where $(a+2b)P_m<aP^{(0)}_x$.  For any choice of $b$, $P_m$ and $P^{(0)}_x$, a sufficiently large value of $a$ always satisfies this constraint.  This means that a beam of neutrinos can always split if the group velocity exceeds the phase velocity, but the splitting rate may be highly suppressed if the phase velocity is only exceeded for a small portion of the set of momenta beneath $P^{(0)}_x$.

\subsection{Splitting rate}
Below we will calculate the splitting rate precisely in a particular model.  In this subsection we will provide a rough estimate of the splitting rate for neutrinos.  The rate is quite similar to that of the decay rate calculated by Cohen and Glashow \cite{cg}, as the relevant process is virtually identical, one need only replace the electron-positron pair with a neutrino-antineutrino pair.  There are still two vertices, leading to a factor of $G_F$ in the amplitude and so $G_F^2$ in the decay rate.  

As we consider more general dispersion relations, there is not necessarily any limiting velocity and so the phase space computation is in general more complicated.  There are still 3 final particles, which have 12 4-momentum components.  The conservation of 4 momentum yields 4 constraints and the mass shell conditions on the 3 particles yield 3 more, leaving 12-4-3=5 undetermined quantities.  Thus the phase space for the decay is 5-dimensional.  Two of these five variables describe the fractional distribution of momenta between the 3 final particles and 3 describe the orthogonal momenta.  The orthogonal momenta, as we have described, would be constrained to vanish if the dispersion relation were linear.  Thus these 3 variables live inside of a region whose radius is proportional to the extra phase space provided by the nonlinearity of the dispersion relations.  

The size of this extra phase space depends upon the choice of neutrino dispersion relation.  For example,  suppose one is interested in splitting at high energies and if at these high energies the dispersion relation is effectively $E=P+aP^k$ with $k>1$, where $a$ has energy dimension $[1-k]$.  Furthermore consider energies low enough so that $aP^k<<P$, so the relativistic term is still dominant.  Now the extra energy available is proportional to $aE^k$ and so the phase space contribution to the decay rate is proportional to $a^3$.

Putting these results together, the decay rate is proportional to $G_F^2 a^3$ which has energy dimension $[-1-3k]$.  The decay rate should have units of energy, therefore by dimensional analysis the decay rate is proportional to
\beq
\Gamma\sim G_F^2 a^3 E^{2+3k}.
\eeq

\section{Splitting and OPERA}

In the remainder of this note we will illustrate the power of neutrino splitting constraints by using them to demonstrate that the neutrino velocity anomaly could not have been caused by any modified dispersion relation for the neutrino alone, in other words a model consistent with OPERA's claim would have to modify the dispersion relations of other particles, which would be very difficult to achieve consistently with experimental constraints.

Following Refs. \cite{ether,me} we will consider dispersion relations $E(P,\rho)$ which are general functions of the norm $P$ of the 3-momentum and background density $\rho$.   An arbitrary dependence on the density $\rho$ is included in order to avoid friction with bounds on neutrino superluminalities from SN1987A \cite{snneut,snphot}, which have plagued previous attempts at the determination of such dispersion relations \cite{miao,dispersione,piudispersione}.   The SN1987A bound may be satisfied by chosing the $\rho$-dependence such that at small $\rho$ neutrinos are subluminal.  Alternately the density-independent case, which includes for example the models of Refs. \cite{ealfa,strumia} is included in our analysis as the special case in which the $\rho$ dependence is trivial.  In what follows we will make no assumptions concerning the $\rho$-dependence, our arguments will be completely general.  However we will assert that the dispersion relations of all particles except for neutrinos are as in the Standard Model.

%The group velocity of a neutrino is determined by its dispersion relation.  In this note, we ask whether OPERA's superluminal neutrino announcement \cite{opera} can be consistently explained \cite{ether,me} by any dispersion relation $E(P,\rho)$ for the neutrino's energy $E$ as a function of the norm of its 3-momentum $P$ and the background density $\rho$ if nothing else is changed in the Standard Model. 

To reduce the number of dispersion relations that we must consider, we use two tricks.  First of all, we restrict our attention to experiments inside of the Earth's crust.  Here the density is more or less constant.  In a region of constant density $\rho_0$, the density-dependence of the dispersion relation is irrelevant and so for these experiments one need only consider a function in one variable $E(P)=E(P,\rho_0)$.  Second, we impose a kind of monotonicity condition on $E(P)$.  In practice we impose that it has at most one inflection point by expanding to second order in $P$ within a given interval, however this assumption can be weakened to allow as many as 10 inflection points without changing our conclusions.  If, on the contrary, there were numerous small windows in $P$ at which the neutrino is extremely subluminal, but these windows were so small than no neutrino yet observed by OPERA lies inside of one, then our conclusions would be evaded.

With these restrictions on the function $E(P)$ we apply three experimental constraints. The first constraint arises from the energies of the neutrinos observed in the 2008 and 2009 OPERA runs and reported in Ref.~\cite{operafeb}.  This spectrum has two features which will be relevant here.  First of all, it has a very long high energy tail, with 4 percent of events above 100 GeV.  Second of all, it fit expectations from simulations remarkably, even at energies of up to 100 GeV.  The high energies are relevant because the effects which will be considered are much stronger at high energy, where more phase space is available, and so they imply stronger bounds.  This constrains the energy loss of the neutrinos during propagation, which places constraints on neutrino splitting and therefore as we will see on the excess of the group velocity over the phase velocity.

The second experimental constraint comes from OPERA's short extraction run, which was able to unambiguously determine, up to whatever systematic errors there may be, the anticipated arrival times of 20 neutrinos \cite{opera}.  All of them arrived earlier than they would have had they traveled at the speed of light.  Their velocities showed fractional excesses $\epsilon=(v-c)/c$ of between $1.7$ and $3.9\times 10^{-5}$.  The energies of these neutrinos are not  known, indeed 14 interacted outside of the detector.  Nonetheless, they can be determined statistically using the energy distribution reported in Ref.~\cite{operafeb}.  The velocities determine the derivative of the dispersion relation in the momentum regime probed by OPERA.

The final constraint that we will use is from the ICARUS experiment, whose detector is subjected to the same neutrinos as its cavemate OPERA.  It did not observe any high energy, nearly collinear, $e^-e^+$ pairs \cite{icarus}.  The resulting bound can be strengthened and extended to higher energies by results from the NOMAD experiment \cite{nomad} and ICECUBE \cite{icecube}, but there is not yet evidence for neutrino superluminality at these energy scales and so this will not strengthen our claim.  Generalizing Cohen and Glashow's results \cite{cg} for this process to a more general dispersion relation, this places a very strong bound on $E(P)-P$.

These three constraints on the neutrino dispersion relation at OPERA energies\footnote{We do not have these bounds at lower energies, but we do not need them.  We consider only neutrino splitting in which all of the resulting neutrinos are within this energy range.   The inclusion of decays with products below this energy range could only increase the decay rate, and so strengthen our result.} force it, in this energy range, to respectively not be too concave $E\p(P) \lesssim E(P)/P$ (neutrino splitting), to have a slope of $1+\epsilon$ (neutrino superluminality) and to approximately lie in the triangle $E(P)\lesssim P$ (Cohen-Glashow Bremsstrahlung). The purpose of this note is to make these bounds quantitative, and so to argue that no such function exists even if the bounds are chosen very conservatively.  Therefore a model which explains OPERA's superluminality claim must do more than just modify the dispersion relation of the neutrino.  In the conclusion we will briefly mention some such additional features that have been considered in the literature which may render such a model consistent.

In Sec.~\ref{vincoli} we will discuss the three experimental constraints that we will need to derive this argument.   To derive the most robust bound, we will only assume neutrino superluminality in the energy regime in which it may be inferred from OPERA data.  In Sec.~\ref{vel} we will determine this regime and so use the bounds to determine the features of a dispersion relation which are necessary to satisfy the three constraints.  We will then consider the dispersion relation which minimizes the neutrino splitting probability, and show in Sec.~\ref{numeri} that nonetheless it yields so much splitting as to be incompatible with OPERA's results. 

\section{The Three Constraints} \label{vincoli}

There are many functions of the form $E(P,\rho)$.  We will greatly simplify the analysis by using only constraints at a fixed value of $\rho$.  More precisely, we will consider only experiments in the Earth's crust, approximating its density to be constant.  In particular, as most of the travel time of neutrinos from SN1987A was in much lower density media, we will not use these to fix any constraint.  Fixing $\rho$, we arrive at a function only of momentum $E_\rho(P)$.  As we will be interested in experiments in which the momenta are roughly parallel to the surface of the Earth, we will ignore the potential directional dependence and only consider the dependence on the magnitude of the momentum.  This leaves us with a function $E(P)$ of a single variable.   In this section we will describe the three experimental constraints which $E(P)$ must satisfy if no new physics is introduced apart from the modification of the neutrino dispersion relation.

\subsection{Neutrino superluminality} \label{supersottosez}

It is most likely that neutrinos are never superluminal.  It is now clear that the superluminality claimed by OPERA was due to experimental error \cite{zichichi}.  However, in this demonstration of the use of neutrino splitting constraints we are attempting to understand general features of models which explain OPERA's neutrino superluminality claim.  Therefore we will impose that neutrinos are superluminal in the region suggested by OPERA.  This is certainly not a well-defined mission.  It may well be that the neutrino velocity is a function of energy which is reasonably constant but has very deep subluminal troughs which are so thin that no OPERA neutrinos have hit them, and so they have not been observed.  The existence of such troughs is very difficult to falsify, as they can be made as thin as one likes, and can in fact lead to a consistent dispersion relation, although one would need about 10 troughs in order to do this.  Therefore we will assume that no such deep, narrow troughs in the neutrino velocity as a function of energy exist.  For simplicity we will chose a constant velocity as a function of energy in the regime probed by OPERA's short extraction run.  It will be clear that our result holds even if this constant is the minimum value allowed by the data, or even two or three standard deviations less.  However we stress that our result is not qualitatively changed by any energy dependence of the dispersion relation, so long as the velocity function does not include at least 10 deep and narrow troughs.

The group velocity of a neutrino is given by the usual formula from classical wave mechanics
\beq
v_g=\frac{\partial E(P)}{\partial P}
\eeq
and so this superluminal velocity corresponds to the slope of $E(P)$, which is slightly greater than unity, throughout the regime probed by this second run.

\subsection{Cohen-Glashow Bremsstrahlung and ICARUS} \label{cgsottosez}

As explained above, if a neutrino requires more energy to arrive at a given momentum than two electrons, then phase space will be available for the neutrino to emit $e^-$ $e^+$ pairs.  Each such emission reduces the neutrino energy by a factor of order 1 \cite{cg}, and so if most neutrinos in any energy range experience such an emission, the spectrum will be distorted significantly.  The strongest bound however comes from the high energy $e^-$ $e^+$ pairs themselves.  The linear dispersion relation considered by Cohen and Glashow would have resulted in millions of such pairs being detected by ICARUS, but in fact none were seen \cite{icarus}.  This fact was used by ICARUS to derive a bound on the neutrino superluminality.  They assumed the linear dispersion relation of Cohen and Glashow, however this derivation may be applied individually to neutrinos in any given momentum range, it does not depend\footnote{Strictly speaking, this is only true for decays in which all momentum is transfered into the electrons, in which case beta decay constraints on the neutrino dispersion relation yield a sufficiently low energy to justify our approximation.  This can be extended to a sufficient fraction of phase space so long as superluminality stays below the 1 percent level.  More generally, the phase space for this decay is dominated by values of $P\p$ in the OPERA energy range.  In this regime the slope of $E$ is greater than unity which means that the Cohen-Glashow process is kinematically allowed and likely sufficient alone to exclude these models.} upon the relation to the function $E(P)$ at other values of $P$.  Therefore, the conservative bound that they give of
\beq
\epsilon=\frac{v-c}{c}<10^{-8}
\eeq
for neutrinos near 20 GeV may straightforwardly be applied to the regime of neutrinos between 40-60 GeV.  In fact, the real bound in this regime is stronger, due to the higher energies which far outweigh the fact that one is only using about 10 percent of the total neutrinos in deriving the bound.  But this will be more than sufficient.

Summarizing, we arrive at our second constraint.  For neutrino momenta in the range of 40-60 MeV,  $E<(1+10^{-8})P$.

\subsection{Neutrino splitting}

The constraints from superluminality and from ICARUS may well be reconciled by a model in which neutrinos are subluminal at low energies and superluminal at high energies \cite{dispersione}.  In such models, $E(P)$ stays below the diagonal line $E<P$ by at first having a slope of less than unity below OPERA energies, falling below the diagonal, and having a slope greater than unity at OPERA energies without ever climbing back to $E=P$.  Before reaching $E=P$, above the energies probed by OPERA, the superluminality can be turned off with $E<P$ at least up to 400 GeV, so as to be consistent with NOMAD.  The region between 1 GeV and 10 GeV is constrained by MINOS \cite{minos}, which does not convincingly demonstrate superluminal neutrinos but does convincingly demonstrate that neutrinos are not too subluminal.  However, one can always compensate for this constraint by forcing neutrinos to be ever more subluminal at low energies.  The only important constraint is that $E(0)$ be smaller than 1 eV, so as to be consistent with bounds from beta decay.  Thus a very concave $E(P)$ may explain OPERA's observed neutrino velocity in a way which is compatible with ICARUS' lack of observed high energy electron positron pairs.

The problem with this scenario is neutrino splitting \cite{splitvec}, which as has been emphasized in Refs.~\cite{splitnuov} strongly constrains the function $E(P)$.  The constraints in those papers do not directly apply here, as we are allowing functions of $\rho$ which can easily accommodate constraints from SN1987A, and we are not assuming any particular form of the dispersion relations.  However, we will see that the applicable constraints are sufficient for our goals.

In this process a neutrino with an initial momentum $P$ and energy $E(P)$ decays into 2 neutrinos and 1 antineutrino with a total momentum $P$ and total energy $E(P)$.  This is possible kinematically only if the function $E(P)$ is concave or more precisely, as we saw in Sec.~\ref{splitsez}, if the group velocity exceeds the phase velocity
\beq
\frac{\partial E(P)}{\partial P}>\frac{E}{P}
\eeq
 in a large enough window of momenta beneath that of the incoming neutrino.

Once the splitting process is allowed kinematically, it will slow down the neutrinos.  The neutrinos lose an order one fraction of their momenta in each reaction.  Like the Cohen-Glashow process, this will ruin the agreement between the neutrino energy spectra and that predicted by simulations in Ref.~\cite{operafeb}.  In particular it will lead to a maximum neutrino energy, but recall that we will only impose superluminality in the energy range in which OPERA has measured neutrino velocities in its short extraction run, and so we only have access to the dispersion relations in this regime.  Therefore this is only in contradiction with OPERA if the energy spectrum is cut off inside of this energy regime, whereas a cutoff in the long high energy tail cannot be predicted.  Finally, each of these processes will cause some deflections to the neutrino beam, making it less likely to hit the OPERA detector and therefore reducing the flux, again leading to tension with Ref.~\cite{operafeb}.  In Sec.~\ref{numeri} we will calculate the corresponding decay rate for OPERA neutrinos for dispersion relations consistent with constraints from neutrino velocity and ICARUS, and we will find that it is dramatically inconsistent with the observed neutrino spectrum.

%\section{Neutrino Splitting}\label{numeri}

\section{OPERA's measurement of the neutrino velocity} \label{vel}

OPERA's neutrinos are sent to Italy from the CNGS experiment at CERN.  Most of them are the decay products of pions, which are created when 400 GeV protons ejected from the SPS in France hit a graphite target in Switzerland.  The fact that the original protons have about 400 GeV each suggests that the neutrino energy distribution will have a tail extending to about 400 GeV.  As the effects discussed in this note become more important at higher energy, for example with Cohen and Glashow predicting an energy loss rate proportional to $E^6$ \cite{cg}, it seems likely that future OPERA runs can lead to bounds which are much tighter than those in this note.  This in particular applies to processes in which neutrinos emit $\mu$ $\bar{\mu}$ pairs, $\pi^0$'s and $\gamma$'s.  

However during OPERA's 2009, 2010 and 2011 runs, due to the fact that it was using CNGS neutrino extractions which were 20 times longer than the effect for which it was searching, velocity measurements were statistical and indirect, and so could give rather little information about the energy dependence on the neutrino velocity.  The results were most consistent with an energy independent distribution, but within two standard deviations could be fit by almost any distribution.  

The statistical error due to the long neutrino extraction time was essentially eliminated when OPERA received short pulses between October 22nd and November 6th of this year, allowing it to determine the velocities of individual neutrinos.  However, given the short time dedicated to this experiment before SPS needed to begin accelerating lead ions, only 20 acceptable events were produced.  These events all yielded measured fractional superluminalities between $1.7$ and $3.9\times 10^{-5}$.  This was interpreted as a more than 6 $\sigma$ signal for neutrino superluminality, but what does it tell us about the energy dependence of this superluminality?

%\begin{figure}%[!htp]
%\begin{center}
%\includegraphics[width=0.75\linewidth]{opera-energia.eps}
%\caption{2008 and 2009 OPERA events with at least one identified muon, compared to predictions from a simulation.  Taken %from Ref.~\cite{operafeb}.}
%\label{energia}
%\end{center}
%\end{figure}

To determine this, first of all, one needs to consider the energy distribution of the neutrinos seen by OPERA.  This is shown in Fig.~10 of Ref.~\cite{operafeb}, compared with the expected results from a simulation.  As one can see, in all energy regimes the observed neutrino spectrum is very close to that predicted, therefore effects like neutrino splitting and Cohen and Glashow's electron-positron pair production must be quite small throughout this range.  In particular there is a long tail at high energies.  While only 0.6\% of the neutrinos from CNGS have energies about 100 GeV, the fact that higher energy neutrinos interact more strongly means that these account for 4\% of all OPERA neutrinos \cite{operafeb}.

In what energy range do the 20 observed events lie?  It is impossible to know, since 14 events occurred outside of the detector.  However, based on the distribution in Fig.~10 of Ref.~\cite{operafeb}, one can determine statistically where they are likely to have occurred.  As we are only interested in establishing a bound, we can safely use a very conservative estimate, and state that they all occurred between 10 and 50 GeV.   What does the distribution of time delays then tell us about the energy dependence of the fractional superluminality $\epsilon=(v-c)/c$?  Again, an energy independent value is favored, but a linear or reciprocal relation cannot be excluded.  For simplicity we will use a constant distribution at the favored value
\beq
\epsilon=2.5\times 10^{-5}.
\eeq
However we note that the splitting rate is proportional to $\epsilon^3$, and so even if we used a constant value as disfavored as $10^{-5}$, this would only decrease the splitting rate by a factor of 20 which we will see is not sufficient to establish consistency.  Moreover, a function which oscillates but rests within this regime, such as a linear or quadratic momentum-dependent fit, would lead to more splitting than a constant choice at the minimum value $10^{-5}$, therefore our analysis will also apply to these more general polynomial dependences.

Now we are ready to assemble the superluminality constraint from Subsec.~\ref{supersottosez} with the ICARUS bound from Subsec.~\ref{cgsottosez}.  The former states that for $10\ {\rm{GeV}}<P<50\ {\rm{GeV}}$ the slope is $1+2.5\times 10^{-5}$.  The second states that during this entire interval it lies below the diagonal $E=P$ or if it surpasses the diagonal, it only surpasses by $10^{-8}P$.  Of course this $10^{-8}P$ is irrelevant, if indeed $E(50\ \rm{GeV})=(1+10^{-8}) 50$\ GeV, then $E(49.98\ \rm{GeV})=49.98$\ GeV and so one may simply reduce the upper limit on $P$ from 50 GeV to 49.98 GeV and conclude that $E<P$ everywhere.  For simplicity we will state
\beq
E(P)<P \hspace{1cm}\rm{In\ the\ range\ } 10\rm{\ GeV}<P<50\rm{\ GeV}.
\eeq

\begin{figure}[!htp]
\begin{center}
\includegraphics[width=13cm,height=7cm]{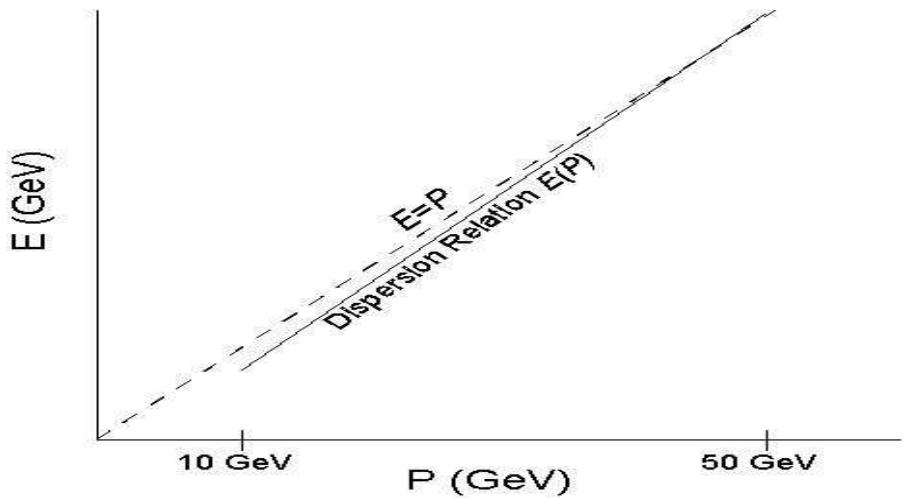}
\caption{The dispersion relation.  The velocity measurement at OPERA implies a slope of between $1+1.7\times 10^{-5}$ and $1+3.9\times 10^{-5}$ when 10 GeV$<P<$50 GeV.}
\label{disper}
\end{center}
\end{figure}

So what happens for $P<10$\ GeV?  It does not matter.  We can consider neutrino splittings in which all three products have more than 10 GeV.  The true rate, including all neutrinos, can only be larger, and so this establishes a lower bound on the decay rate.  The decay rate of a 50 GeV neutrino to 3 neutrinos at energies between 10 GeV and 30 GeV can be determined using the dispersion relation described here.  Notice that this dispersion relation depends on only two things.  First, its slope is determined by superluminality which is measured directly by the OPERA experiment.  Next, its height receives only an upper bound, the $E=P$ line given by the constraint from ICARUS.  Therefore in principle we have not determined the dispersion relation, it may always be shifted to lower energies.  However such a shift will only make the dispersion relation more concave, and so increase the neutrino splitting decay rate.  The most conservative calculation of the decay rate corresponds to the case $E(50\ \rm{GeV})=50$\ GeV.  In the next subsection we will determine the decay rate for this case and see that indeed neutrinos almost never arrive at OPERA from CNGS without splitting, in grave contradiction with Fig.~10 of Ref.~\cite{operafeb}. 

\section{Splitting decay rate} \label{numeri}

We assume that in the range between 10 and 50 GeV the neutrinos can be described by the effective dispersion relation in Fig. \ref{disper}
\beq\label{dispRel}
E(P)=(1+\epsilon)(P-B)
\eeq
where
\beq\label{const}
\epsilon\simeq2.5\times10^{-5}, \qquad B=\frac{\epsilon}{1+\epsilon}50\textrm{ GeV}\simeq 1.2 \textrm{\ MeV} .
\eeq

This is consistent with observations, indeed
\begin{enumerate}
\item In the range between 10 and 50 GeV, the group velocity is
\beq
|v_i|=\left|\frac{\partial E}{\partial p_i}\right|=1+2.5 \times 10^{-5} .
\eeq
\item At 50 GeV the neutrino 4-momentum  is light-like (\ie $E(50\ \textrm{GeV})=50\ \textrm{GeV}$). 
\end{enumerate} 
Finally, we have set the neutrino masses to zero, since they are negligible at these energies.

The Lorentz violating kinetic term that can reproduce the effective dispersion relation  (\ref{dispRel}) is
\beq \label{Lag}
\mathcal{L}_k=\bar{\nu}(p_0+B)\gamma^0\nu+(1+\epsilon)\bar{\nu}p_i\gamma^i\nu
\eeq
where $\epsilon$, $B$ are the constants defined in (\ref{const}).

\subsection{Notation}
We are interested in the splitting process
\beq
\nu(p)\rightarrow \nu(p')+\nu(k)+\bar{\nu}(k').
\eeq
Let us briefly clarify our notation: we use lowercase letters for the 4-momenta $p$, $p'$, $k$ and $k'$ and uppercase letters for the magnitudes of the spatial vectors: $P=|\vec{p}|$.  The scalar product between two  4-vectors as $p$ and $k$ will be denoted by $p\cdot k$, while if only the spatial components of the momentum are involved we will  write $\vec{p}\cdot \vec{k}$. 
$E_p$ will denote the zero component of the 4-momentum p and so $p\cdot p=E_p^2-|\vec{p}|^2=E_p^2-P^2$.

%Summarizing:
%\begin{itemize}
%\item $p$ can indicate both the quadri-momentum $p_\mu$ and the magnitude of the three-%vector $\vec{p}$
%\item 
%\item $p\cdot k$ denotes the scalar product between two quadri-vectors, while the same %operation between two three-vectors is indicated with $\vec{p}\cdot \vec{k}$ (so, for example %$p\cdot p=E_p^2-|\vec{p}|^2=E_p^2-p^2$)
%\end{itemize}

\subsection{Cross Section}
We will now calculate the decay rate for our neutral current process, neutrino splitting. The coupling between the neutrinos and the $Z$ gauge boson is
\beq
\frac{g}{2\Cos\theta_W}=2^{1/4}G_F^{1/2}m_Z .
\eeq
The multiplicity factor of the relevant Feynman diagram is 2, and the corresponding amplitude is given by
\beq
\sum_{spin}\left|\mathcal{M}\right|^2=\sum_{spin}8 G_F^2\left|\left(\bar{u}(p)\gamma_\mu u(p')\right)\left(\bar{u}(k)\gamma^\mu v(k')\right)\right|^2 .
\eeq
In the Lorentz-violating theory in which the kinetic term is given by the (\ref{Lag}), the completeness relations are
\beq
\sum_{spin} u^s(p)\bar{u}^s(p)=v^s(p)\bar{v}^s(p)=\widetilde{p}\!\!\!/
\eeq
where
\beq
\widetilde{p}=(p_0+B,(1+\epsilon)p_i) .
\eeq
Similar notation has been introduced in Refs. \cite{dirac,tildearticolo} in this context.  

After some straightforward algebraic manipulations, we find
\beq
\sum_{spin}\left|\mathcal{M}\right|^2=128 G_F^2(\widetilde{p}\cdot\widetilde{k'})(\widetilde{p'}\cdot\widetilde{k})
\eeq
and the decay width
\bea
\Gamma&=&\frac{1}{2E_p}\int\frac{\textrm{d}^3\vec{p'}}{(2\pi)^32E_{p'}}\int\frac{\textrm{d}^3\vec{k}}{(2\pi)^32E_{k}}\int\frac{\textrm{d}^3\vec{k'}}{(2\pi)^32E_{k'}}\frac{1}{2}|\mathcal{M}|^2(2\pi)^4\delta^4(p-p'-k-k')\nonumber\\ 
&=&\frac{8G_F^2}{(2\pi)^5 E_p}
\int\frac{\textrm{d}^3\vec{p'}}{E_{p'}}
\int\frac{\textrm{d}^3\vec{k}}{E_{k}}
\int \frac{\textrm{d}^3\vec{k}}{2E_{k'}} (\widetilde{p}\cdot\widetilde{k'})(\widetilde{p'}\cdot\widetilde{k})\delta^4(p-p'-k-k') =\frac{8G_F^2}{(2\pi)^5 E_p}\Gamma_{r}.\nonumber\\
\eea

We will restrict the domain of integration to energies between 10 and 50 GeV, since OPERA only indicates that (\ref{dispRel}) holds  in this range. This means that the value $\Gamma$ that we will obtain is only a lower bound; however we will see that it is sufficient for our purposes.  More or less following the phase space integration strategy of Ref. \cite{miao} we find
\bea
\Gamma_{r}&=&
\int\frac{\textrm{d}^3\vec{p'}}{E_{p'}}
\int\frac{\textrm{d}^3\vec{k}}{E_{k}} \int d^4k  (\widetilde{p}\cdot\widetilde{k'})(\widetilde{p'}\cdot\widetilde{k})
\left.\delta(\widetilde{k'}\cdot \widetilde{k'})\right|_{E_{k'}>10}\delta^4(p-p'-k-k')\label{GamR}\\
&=& \int\frac{d^3\vec{p'}}{E_{p'}}
\int\frac{K^2\textrm{d}K\textrm{d}\theta_1\sin\theta_1\textrm{d}\psi_1}{E_{k}} (\widetilde{p}\cdot(\widetilde{p}-\widetilde{p'}-\widetilde{k}))(\widetilde{p'}\cdot\widetilde{k})
\left.\delta(|\widetilde{p}-\widetilde{p'}-\widetilde{k} |^2)\right|_{E_{p}-E_{p'}-E_k>10}\nonumber
\eea
where 
$|\widetilde{p}-\widetilde{p'}-\widetilde{k} |^2=(\widetilde{p}-\widetilde{p'}-\widetilde{k}) \cdot (\widetilde{p}-\widetilde{p'}-\widetilde{k})$.

It is convenient now to define $\theta_1$ as the angle between $\vec{k}$ and $\vec{p}-\vec{p'}$ and perform the change of variables $x=\cos\theta_1$. Similarly, we let $\theta_2$ be the angle between $\vec{p}$ and $\vec{p'}$ and $y=\cos\theta_2$. We will need the relations
\beq
\vec{p}\cdot\vec{k}=PK\frac{(P-P\p y)x+P\p\sqrt{1-x^2}\sqrt{1-y^2}\cos\psi_1}{|\vec{p}-\vec{p\p}|}
\eeq
and
\beq
\vec{p'}\cdot\vec{k}=\vec{p}\cdot\vec{k}-(\vec{p}-\vec{p'})\cdot\vec{k}=\vec{p}\cdot\vec{k}-|\vec{p}-\vec{p\p}|Kx.
\eeq
We can write the last delta function as $\delta(f(x))$ where the function f(x) has a single root
\beq
x_0=\frac{|p-p\p|^2-(P-P\p+2B)^2+2k(P-P\p+2B)}{2K|\vec{p}-\vec{p'}|}
\eeq
and its first derivative reads
\beq
f'(x)=2(1+\epsilon)^2k|\vec{p}-\vec{p'}|.
\eeq
After integrating over $x$, Eq. (\ref{GamR}) reads
\beq
\Gamma_{r}= \int\frac{d^3\vec{p'}}{E_{p'}}
\int\frac{K\textrm{d}K\textrm{d}\psi_1}{E_{k}}\left. \frac{(\widetilde{p}\cdot(\widetilde{p}-\widetilde{p'}-\widetilde{k}))(\widetilde{p'}\cdot\widetilde{k})}{2|\vec{p}-\vec{p'}|}
\right|_{E_{p}-E_{p'}-E_k>10,x=x_0}.\nonumber
\eeq

The domain of integration is determined by the conditions $-1 \leq x_0 \leq 1$. Moreover, our phase space is limited to the region in which all the neutrinos have energy between 10 and 50 GeV, yielding the additional conditions
\bea
&& P-P\p-K>10\rm{\ GeV} \label{cond1}\\
&& K>10\rm{\ GeV}. \label{cond2}
\eea
Notice that in this case we have neglected the contributions to the energy proportional to $\epsilon$. 
Indeed, following the same argument used in the previous section, one can easily see that they are equivalent to an irrelevant shift of the region of validity of our effective dispersion relation.

From the conditions on $x_0$ we obtain
\bea
&&\frac{P-P\p+2B-|\vec{p}-\vec{p\p}|}{2}\leq K \leq \frac{P-P\p+2B+|\vec{p}-\vec{p\p}|}{2}\label{cond3}\\
&&y>1-\frac{2B(B+P-P\p)}{P\p P}.\label{cond4}
\eea
If the condition (\ref{cond4}) is not satisfied, $x_0\leq1$ implies
\beq 
 K \geq \frac{P-P\p+2B+|\vec{p}-\vec{p}'|}{2}
\eeq
which is in contradiction with the condition (\ref{cond1}). Since it easy to see that (\ref{cond1}) and (\ref{cond2}) are stronger bounds than (\ref{cond3}) our domain of integration is given by
\bea
&&10\rm{\ GeV}\leq K \leq P-P\p-10\rm{\ GeV}\label{conD1}\\
&&y>1-\frac{2B(B+P-P\p)}{P\p P}.
\eea
The conditions (\ref{conD1}) also imply that $10\rm{\ GeV}<P-P\p-10\rm{\ GeV}\Rightarrow P\p<P-20\rm{\ GeV}$, which yields the last bound.
Choosing $P=50$\ GeV and performing the integration we obtain the decay rate
\beq
\Gamma=3.8 \times 10^{-19}\textrm{ GeV}
\eeq
and the decay length 
\beq
\ell=\frac{\hbar c}{\Gamma}=0.52\textrm{ km} .
\eeq

This is much shorter than the observed 730 km distance that neutrinos travel from CNGS to OPERA.  Therefore such models are strongly excluded, because the observed neutrino flux would not fit simulations as well as seen in Fig.~10 of Ref.~\cite{operafeb}.  As in the case of Ref. \cite{cg} the decay rate at a fixed energy is proportional to the cube of the difference between the electron and neutrino dispersion relations, therefore a suppression of this process requires that these two energies be at least an order of magnitude closer.  As the slope at most momenta $P$ is fixed directly by the OPERA velocity measurement, this requires that the neutrino energy curve in Fig. \ref{disper} by modified by including narrow corrections in which the neutrino energy rapidly approaches the electron energy, corresponding to a massively subluminal neutrino in a very narrow momentum band.  To reduce the difference between the two dispersion relations by an order of magnitude, one requires at least 10 such narrow bands.  In a polynomial fit to the dispersion relation, this requires a polynomial of order much greater than 10.

\section{Conclusion}

In this note we have seen that the neutrino splitting process $\nu\rightarrow 2\nu+\bar{\nu}$ places strong restrictions on the form of the neutrino dispersion relation $E(P)$.  The result is effectively that $E(P)$ will be linear or convex, or more precisely that the group velocity will never exceed the phase velocity.  This is a quantum analogue of the modulational instability in classical wavemechanics\footnote{For a review of various manifestations of these instabilities in classical physics, see for example Ref.~\cite{mod}.} in which nonlinearities (corresponding to the self-interactions in the quantum theory) cause a plane wave to decompose into pulses.  As in the quantum case, this occurs when nonlinearities in the dispersion relation yield a group velocity greater than the phase velocity in a certain range of the space of momenta.   However in the classical case the number of excitations does not change, the instability serves only to drive the Fourier components of the waveform away from the original planewave momentum.

\subsection{Applications for neutrino splitting}

%\footnote{Such a stability condition can already be seen in classical wave mechanics.  The classical analogue of the splitting process is the modulational instability in anomalous dispersive media.}.  
As the splitting rate depends only upon the neutrino sector, it may be used to place robust constraints on a theory, independently of the dispersion relations of the other particles.   It therefore has numerous applications.  In this note we have seen that it implies that a change to the neutrino dispersion relation alone, even with an arbitrary density dependence, is not sufficient for consistency with OPERA and ICARUS results.  %There are many proposed methods of avoiding this problem, by changing aspects of the theory apart from the neutrino dispersion relation.  For example, if the electron and neutrino dispersion relations are equal at high energies, as indeed is suggested by SU(2) gauge-invariance, then the Cohen-Glashow process is kinematically forbidden.  Therefore the condition that $E=P$ at 50 GeV is relaxed and the above inconsistency is lifted.  However, in the 50-100 GeV range one needs to consider processes like $\nu\rightarrow\nu+\pi^0$, $\nu\rightarrow\nu+\gamma$ and $\nu\rightarrow\nu+\mu+\bar{\mu}$, which are not eliminated by a modification of the electron dispersion relation.  If OPERA continues to pursue short pulse runs, it may well be able to extend its superluminality results into this range.  At that point one may be forced to render even more particles superluminal, as in the second paper of Ref.~\cite{me}.  If all of these dispersion relations are linear and identical for all particles, this would be equivalent to a universal change in the effective metric, changing Einstein's equations in matter.  More immediately, it is possible to calculate the constraints from these processes and to see if and when indeed other dispersion relations must be altered.

The concavity bounds may also be used to place upper bounds on superluminality at high energies, for example at ICECUBE.  Indeed, using the fact that 400 TeV neutrinos are depleted by at most a factor of 2 with respect to simulations \cite{icecube}, one may reverse the argument of Ref.~\cite{splitvec} to exclude the $\epsilon\sim E$ and $\epsilon\sim E^2$ behaviors which historically have been expected from effective field theories.
 
%But more important is what can be said about lower energies.  It is often said that upcoming data released from MINOS will confirm or deny OPERA's claim.  Of course, this is unclear as MINOS neutrinos are about 10 times less energetic than OPERA neutrinos, perhaps the superluminality is simply too small to observe at MINOS.  However if OPERA neutrinos are superluminal, and MINOS neutrinos are much less superluminal, then there is a positive second derivative in the dispersion relation between the two energy scales, and so OPERA energy neutrinos may split into MINOS energy neutrinos.  It would be useful to calculate these decay probabilities and so to use OPERA's data to predict a lower bound on neutrino superluminality at MINOS.  In this way, OPERA's claim will become falsifiable at MINOS.  Of course, this assumes that no exotic physics is at play such as many flavors of sterile neutrinos, extradimensional shortcuts and modified momentum conservation.

\subsection{Experimental signatures}

In closing, we would like to mention that if indeed the neutrino dispersion relation does depend on density, but that this is consistent with the above constraints because for example charged lepton dispersion relations have a similar dependence, then there are several kinds of experimental signatures for which one may search.

Besides the neutrino, there are two particles that may travel long distances, of order a kilometer, through solid rock.  These are the $\mu$ and the $\pi^0$.  Millions of them, created by interactions with cosmic rays, have been detected in deep detectors such as MINOS.  If their dispersion relations are modified with respect to those of other particles, one may look for signatures of this change in their decay rates inside of various media or even direct measurements of their travel time from the surface to underground detectors.

However it may be that the dispersion relations of all particles have the same density dependence, then what signature is left?  Recall that in 1979 FermiLab \cite{fermi79} was able to place strong bounds on neutrino superluminality, of order $5\times 10^{-5}$, by racing neutrinos and muons a kilometer through rock.  Significant error was introduced due to uncertainties involving the time lost by a muon traveling through rock.  However this experiment was intrinsically more precise than experiments like OPERA and MINOS because there was no reliance on an external clock, one merely measured the time difference the arrivals of the muons and neutrinos at the same location.  This is why a long baseline was not necessary, one kilometer already gave a precision comparable to MINOS' 734 km experiment 25 years later.

One can do a very similar experiment racing neutrinos passing through rock and neutrinos beside them passing through a tunnel.  If one is lucky enough to receive neutrinos that passed both through the rock and through the tunnel in the same extraction, then one can compare their arrival times and thus measure their relative velocities very accurately.  Such conincidences will be much more frequent at MINOS' near detector after the energy increase of its neutrino beam scheduled to take place in two years' time.

\section* {Acknowledgements}

\noindent JE is supported by the Chinese Academy of Sciences
Fellowship for Young International Scientists grant number
2010Y2JA01. EC, XB and XZ are supported in part by the NSF of
China.  We thank Pengfei Yin for useful discussions.

%%%%%%%%%%%%%%%%%%%%%%%%%%%%%%%%

\end{document}